\makeatletter
\def\input@path{{styles/}{figures/}}
\makeatother

\RequirePackage{silence}
\documentclass[journal,twoside,web]{styles/IEEEtran}

\usepackage{tmi}
\usepackage{amsmath,amssymb,amsfonts}
\usepackage{algorithmic}
\usepackage{graphicx}
\usepackage{textcomp}
\usepackage{bbm}
\usepackage{booktabs}
\usepackage{array}
\usepackage{siunitx}
\usepackage[hidelinks]{hyperref}

\newcommand{\NA}{---}
\newcommand{\numberthis}{\addtocounter{equation}{1}\tag{\theequation}}

\begin{document}

\thispagestyle{empty}
\onecolumn
\noindent
\copyright 2023 IEEE. Personal use of this material is permitted. Permission from IEEE must be obtained for all other uses, in any current or future media, including reprinting/republishing this material for advertising or promotional purposes, creating new collective works, for resale or redistribution to servers or lists, or reuse of any copyrighted component of this work in other works.
\newpage
\twocolumn
\setcounter{page}{1}

\title{Concurrent ischemic lesion age estimation and segmentation of CT brain using a Transformer-based network}
\author{Adam Marcus, Paul Bentley, and Daniel Rueckert, \IEEEmembership{Fellow, IEEE} }

\thanks{Manuscript received July, 2022; revised January, 2023; revised again May, 2023; accepted June, 2023. Date of publication X, 2023; date of current version X, 2023. This work was supported by the UK Research and Innovation: UKRI Center for Doctoral Training in AI for Healthcare under Grant EP/S023283/1 and UK National Institute for Health Research i4i Program under Grant II-LA-0814-20007. For the purpose of open access, the author has applied a Creative Commons Attribution (CC BY) license to any Author Accepted Manuscript version arising. \textit{(Corresponding author: Adam Marcus.)} }
\thanks{A. Marcus is with the Department of Computing and the Division of Brain Sciences, Department of Medicine, Imperial College London, London SW7 2AZ, U.K. (e-mail: adam.marcus11@imperial.ac.uk).}
\thanks{P. Bentley is with the Division of Brain Sciences, Department of Medicine, Imperial College London, London SW7 2AZ, U.K.}
\thanks{D. Rueckert is with the Department of Computing, Imperial College London, London SW7 2AZ, U.K.}

\maketitle

\begin{abstract}
	The cornerstone of stroke care is expedient management that varies depending on the time since stroke onset. Consequently, clinical decision making is centered on accurate knowledge of timing and often requires a radiologist to interpret Computed Tomography (CT) of the brain to confirm the occurrence and age of an event. These tasks are particularly challenging due to the subtle expression of acute ischemic lesions and the dynamic nature of their appearance. Automation efforts have not yet applied deep learning to estimate lesion age and treated these two tasks independently, so, have overlooked their inherent complementary relationship. To leverage this, we propose a novel end-to-end multi-task transformer-based network optimized for concurrent segmentation and age estimation of cerebral ischemic lesions. By utilizing gated positional self-attention and CT-specific data augmentation, the proposed method can capture long-range spatial dependencies while maintaining its ability to be trained from scratch under low-data regimes commonly found in medical imaging. Furthermore, to better combine multiple predictions, we incorporate uncertainty by utilizing quantile loss to facilitate estimating a probability density function of lesion age. The effectiveness of our model is then extensively evaluated on a clinical dataset consisting of 776 CT images from two medical centers. Experimental results demonstrate that our method obtains promising performance, with an area under the curve (AUC) of 0.933 for classifying lesion ages $\leq$4.5 hours compared to 0.858 using a conventional approach, and outperforms task-specific state-of-the-art algorithms.
\end{abstract}


\section{Introduction}
\label{sec:introduction}
\begin{figure}
	\centering
	\includegraphics[width=\columnwidth]{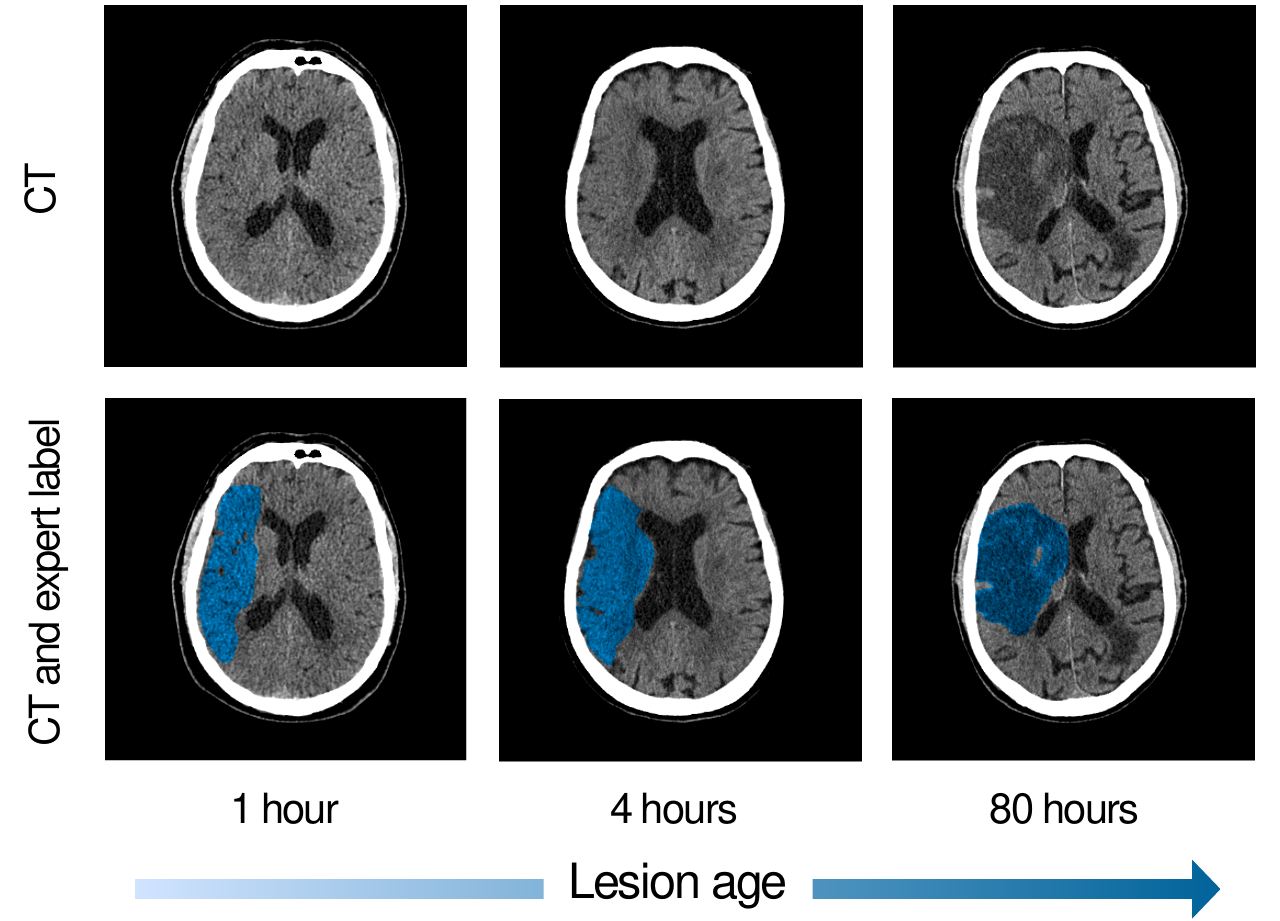}
	\caption{Example images and expert segmentations of different subjects from our clinical dataset illustrating the appearance of ischemic lesions changing over time.}
	\label{fig:lesion-appearance}
\end{figure}

\IEEEPARstart{S}{troke} is the most frequent cause of adult disability and the second commonest cause of death worldwide \cite{who_2018}. The vast majority of strokes are ischemic and result from the blockage of blood flow in a brain artery, often by a blood clot. Consequently, treatment is focused on rapidly restoring blood flow before irrevocable cell death \cite{saver2006time}. The two main approaches are: intravenous thrombolysis, chemically dissolving the blood clot; and endovascular thrombectomy, physically removing the blood clot. Notably, the efficacy of both these treatments decreases over time until their benefit is outweighed by the risk of complications. It is for this reason that current guidelines limit when specific treatments can be given. In the case of thrombolysis, to within 4.5 hours of onset \cite{hacke2008thrombolysis}. Therefore, accurate knowledge of timing is central to the management of stroke. However, a significant number of strokes are unwitnessed, with approximately 25\% occurring during sleep \cite{rimmele2014wake}. In these cases, neuroimaging can help, with previous studies showing promising results using modalities not routinely available to patients, such as magnetic resonance imaging (MRI) and perfusion computed tomography (CT) \cite{ma2019thrombolysis,thomalla2018mri}. Ideally, the widely-available non-contrast CT (NCCT) would be used, but this task is challenging even for detection alone, as early ischemic changes are often not visible to the naked eye (\autoref{fig:lesion-appearance}).

\subsection{Related work}

Several studies have attempted to delineate early ischemic changes on NCCT. The majority have used image processing techniques or machine learning methods based on hand-engineered features but more recent efforts have used deep learning \cite{el2021evaluating}. Qui \textit{et al.} \cite{qiu2020machine} proposed a random forest voxel-wise classifier using features derived from a pre-trained U-Net and achieved a Dice Similarity Coefficient (DSC) \cite{chan2001active} of 34.7\% \cite{kuang2021eis}. Barros \textit{et al.} \cite{barros2020automated} used a convolutional neural network (CNN) and attained a DSC of 37\%. El-Hariri \textit{et al.} \cite{el2021evaluating} implemented a modified nnU-Net and reported DSCs of 37.7\% and 34.6\% compared to two experts. To the best of the authors' knowledge, the current state-of-the-art for this task is EIS-Net \cite{kuang2021eis}, a 3D triplet CNN that achieved a DSC of 44.8\%.

In contrast, few studies have explored using NCCT to estimate the lesion age. Broocks \textit{et al.} \cite{broocks2020lesion} used quantitative net water uptake, originally introduced by Minnerup \textit{et al.} \cite{minnerup2016computed}, to identify patients within the 4.5 hour thrombolysis time window and attained an area under the receiver operator characteristic curve (AUC) of 0.91. Mair \textit{et al.} \cite{mair2021feasibility} introduced the CT-Clock Tool, a linear model using the attenuation ratio between ischemic and normal brain, and achieved an AUC of classifying scans $\leq$4.5 hours of 0.955 with median absolute errors of 0.4, 1.8, 17.2 and 32.2 hours for scans acquired $\leq$3, 3–9, 9–30 and $>$30 hours from stroke onset. These studies all currently require manual selection of the relevant brain regions, and as of yet, have not utilized deep-learning methods that may allow for improved performance.

Deep learning methods have shown great potential across many domains, with convolutional architectures proving highly successful in medical imaging. Here the inductive biases of CNNs, known to increase sample efficiency \cite{d2021convit}, are particularly useful due to the scarcity of medical data. However, this may be at the expense of performance, as Transformers \cite{vaswani2017attention} have surpassed CNNs across many computer vision tasks. By relying on flexible self-attention mechanisms, Transformer-based models can learn global semantic information beneficial to dense prediction tasks like segmentation but typically require vast amounts of training data to do so. Recently, d’Ascoli \textit{et al.} \cite{d2021convit} have attempted to address this by introducing gated positional self-attention (GPSA), a type of self-attention with a learnable gating parameter that controls the attention paid to position versus content and can combine the benefits of both architectures.

Traditionally, Transformer and other deep-learning methods have focused on learning different tasks in isolation with separate networks, yet many real-world problems are naturally multi-modal \cite{vandenhende2020revisiting}. This has contributed to the increasing popularity of multi-task learning (MTL) \cite{caruana1997multitask}, a learning paradigm with the aim of jointly learning related tasks to help improve the generalization performance of all tasks \cite{zhang2021survey}. The underlying insight is that by combining the data of different learning tasks, MTL models can learn robust and universal representations that enable them to be more powerful and reduce the risk of overfitting \cite{zhang2021survey}.

\subsection{Motivation}

As the appearance of an ischemic lesion is highly dynamic, a perfect segmentation model would need to recognize lesions for all time points and implicitly understand the lesion age. These two tasks, lesion segmentation and age estimation, appear to be inherently complementary, and therefore, it seems reasonable that they may benefit from being learned together.

Furthermore, existing work has shown that estimating lesion age can gain from comparing the affected brain to the spatially distant unaffected side \cite{minnerup2016computed}. It may then be particularly advantageous to have a wide receptive field and, consequently, utilizing mechanisms such as a Transformer would seem appropriately suited. However, standard Transformers are typically only effective at large scales, as they lack some of the inductive biases of convolutional neural networks, such as translational equivariance, and thus require large datasets, which are often not available in the medical domain \cite{dosovitskiy2020image}. Hence, it is justifiable to consider architectural modifications and alternative designs of Transformers that have been suggested to address this issue, such as using GPSA modules.

\subsection{Contributions}

In this work, we propose a multi-task network to simultaneously perform the segmentation of ischemic lesions and estimate their age in CT brain imaging. The main contributions are: (1) We introduce a novel end-to-end transformer-based network to solve both lesion age estimation and segmentation. To our knowledge, this is the first time a deep learning-based method has been applied to solve the challenging task of estimating lesion age. (2) We enhance the data efficiency of our approach by integrating GPSA modules into the model and using a CT-specific data augmentation strategy. (3) To further improve the performance of our model at estimating lesion age, we suggest a new method to better combine multiple predictions by incorporating uncertainty through the estimation of probability density functions. The effectiveness of the proposed method is then demonstrated by extensive experiments using CT imaging data from 776 patients across two clinical centers.

\section{Method}

\subsection{Network}

\begin{figure*}
	\centering
	\includegraphics[width=0.8\textwidth]{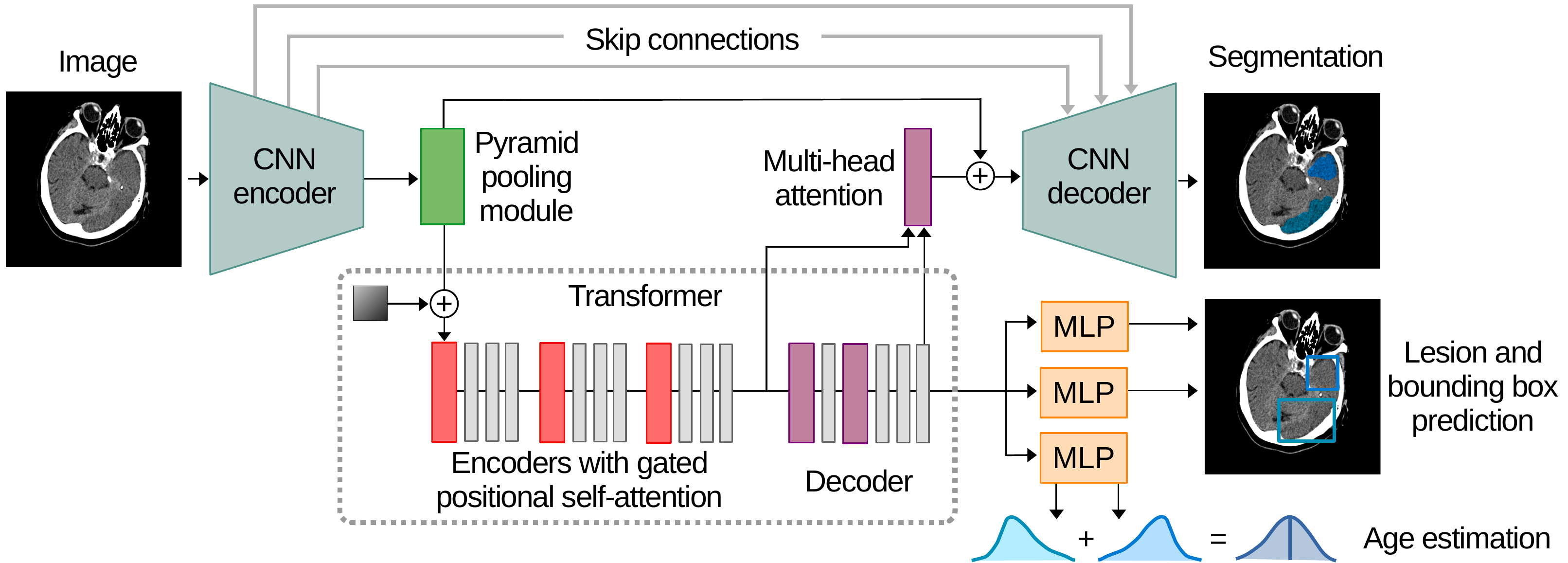}
	\caption{Overview of the proposed model architecture. Input 3D CT images are processed slice by slice. First, a CNN backbone combined with a pyramid pooling module (PPM) extracts image features at multiple scales. Second, a transformer encoder-decoder with gated positional self-attention (GPSA) uses these features to predict output embeddings for several object queries. Third, multi-layer perceptrons (MLP) use these embeddings to predict lesions, bounding boxes, and lesion age probability distributions. Fourth, a segmentation head generates masks for each lesion based on attention. Finally, the per-slice outputs are combined if the predicted masks are connected in 3D, and for each lesion, the most likely age estimate is used.}
	\label{fig:transformer-architecture}
\end{figure*}

An overview of the proposed model is presented in \autoref{fig:transformer-architecture}. The proposed model is based on the DEtection TRansformer (DETR) panoptic segmentation architecture \cite{carion2020end} with modifications to improve sample efficiency, performance, and facilitate lesion age estimation. \autoref{tab:model-differences} summarizes the main architectural differences. All activation functions were changed to the Gaussian error linear unit \cite{hendrycks2016gaussian} (GELU) and batch normalization \cite{ioffe2017batch} was replaced with group normalization \cite{wu2018group} to accommodate smaller batch sizes. The main components of the proposed model are: 1) a CNN backbone; 2) a transformer encoder-decoder; 3) lesion age estimation, and bounding box prediction heads; and 4) a segmentation head.

\begin{table}
	\caption{Summary of main architectural differences between DETR \cite{carion2020end} and the proposed model.}
	\label{tab:model-differences}
	\centering
	\begin{tabular*}{\linewidth}{@{\extracolsep{\fill}}r@{\hskip.2cm}c@{\hskip.2cm}c}
		\toprule
		& Ours & DETR \\
		\midrule
		Activation functions & GELU \cite{hendrycks2016gaussian} & ReLU \cite{nair2010rectified} \\
		Normalization layers & GroupNorm \cite{wu2018group} & BatchNorm \cite{ioffe2017batch} \\
		CNN Backbone & ResNeXt-50 32×4d \cite{xie2017aggregated} & ResNet-50 \cite{he2016deep} \\
		Feature projection & PPM \cite{zhao2017pyramid} & 1×1 convolution \\
		Attention layers & GPSA \cite{d2021convit} & Multi-head attention \cite{vaswani2017attention} \\
		Encoder layers & 3 & 6 \\
		Decoder layers & 1 & 6 \\
		Object queries & 10 & 100 \\
		Prediction heads & BBox, class, and regression & BBox and class \\
		\bottomrule
	\end{tabular*}
	\begin{tabular}{p{\dimexpr\linewidth-2\tabcolsep\relax}}
		\addlinespace[0.5em]
		\begin{footnotesize}
			GELU = Gaussian error linear unit; ReLU = Rectified linear unit; CNN = Convolutional neural network; PPM = Pyramid pooling module; GPSA = Gated positional self-attention; BBox = Bounding box
		\end{footnotesize}
	\end{tabular}
\end{table}

The CNN backbone encoder extracts image features of a 2D CT slice input image. It is comprised of four ResNeXt \cite{xie2017aggregated} blocks and produces an activation map. This activation map is then projected to a feature patch embedding and concatenated with fixed positional encodings \cite{cordonnier2019relationship}. Rather than use a 1×1 convolution as in the original DETR architecture, we use a pyramid pooling module \cite{zhao2017pyramid} (PPM) that has empirically been shown to increase the effective receptive field by incorporating features extracted at multiple scales.

The transformer encoder-decoder learns the attention between image features and predicts output embeddings for each of the $N=10$ object queries. Here $N$ was determined by the maximum number of lesions visible in a given slice. We use three transformer encoder blocks and one transformer decoder block following the standard architecture \cite{vaswani2017attention} with a couple of exceptions. First, rather than using an auto-regressive model \cite{vinyals2015order} we decode the $N$ objects in parallel. Second, to improve the data efficiency of the model we replace the multi-head attention layers in the encoder with GPSA layers.

The lesion, age estimation, and bounding box prediction heads are each multi-layer perceptrons (MLP) and map the output embeddings of the transformer encoder-decoder to lesion, lesion age, and bounding box predictions. These heads process the queries in parallel and share parameters over all queries.

The segmentation head generates binary masks for each object instance based on attention. A two-dimensional multi-head attention layer produces attention heatmaps from the attention between the outputs of the transformer encoder and decoder. These heatmaps are then upscaled by a U-Net \cite{ronneberger2015u} type architecture with long skip connections between the CNN encoder and decoder blocks.

\subsection{Data Augmentation}

\begin{figure}
	\centering
	\includegraphics[width=0.65\columnwidth]{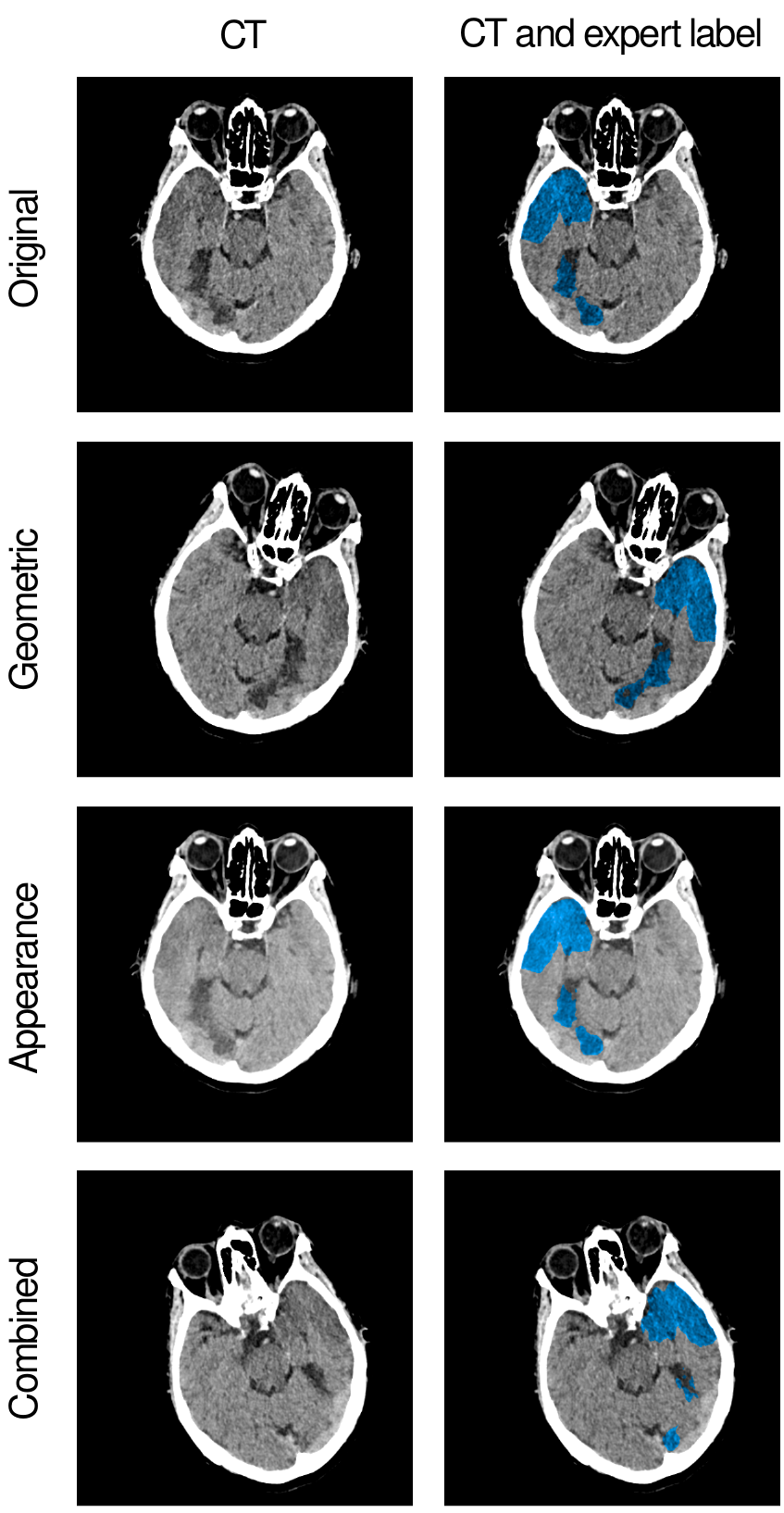}
	\caption{Example CT-specific geometric and appearance transforms.}
	\label{fig:data-augmentation}
\end{figure}

To improve the generalizability of our model and prevent overfitting due to limited training data, we adopted a CT-specific augmentation strategy with geometric and appearance transforms. Geometric transforms included: random axial plane flips; \SI{+-5}{\percent} isotropic scaling; \SI{+-20}{\milli\metre} translation; and \SI{+-0.5}{\radian} axial otherwise \SI{+-0.1}{\radian} plane rotation. Appearance transforms included an intensity transform introduced by Zhou \textit{et al.} \cite{zhou2019models} and a transform we propose to account for the slice thickness variation often present in CT datasets. Regions of the brain are area interpolated \cite{wong1999area} to a random slice thickness, ranging from \SIrange{1}{3}{\milli\metre} to match the sizes in our dataset, then upscaled back to their original shape. Examples of these transforms can be seen in \autoref{fig:data-augmentation}.

\subsection{Loss Function}

We use a combined loss function to enable direct set prediction. The set prediction $\hat{y} = \{\hat{y}_{i} = \{\hat{p}_{i}, \hat{b}_{i}, \hat{s}_{i}, \hat{a}_{i}\}\}_{i=1}^{N}$ consists of the lesion probability $\hat{p}_{i} \in \mathbb{R}^2$ (lesion or no lesion), bounding box $\hat{b}_{i} \in \mathbb{R}^4$, segmentation mask $\hat{s}_{i} \in \mathbb{R}^{H \times W}$ where $H \times W$ is the spatial resolution, and lesion age quantiles $\hat{a}_{i} \in \mathbb{R}^3$ for each of the $N$ object queries. To ensure the loss function is invariant to permutation of the predictions, the Hungarian algorithm \cite{kuhn1955hungarian} was used to assign each instance set label $y_{\sigma(i)}$ to the corresponding query set prediction $\hat{y}_{i}$ where $\sigma_i$ represents the best matching order of labels. The combined loss $\mathcal{L}$ is normalized by the number of lesions in a batch and comprises of a lesion loss $\mathcal{L}_p$, bounding box losses $\mathcal{L}_b$ and $\mathcal{L}_g$, segmentation losses $\mathcal{L}_f$ and $\mathcal{L}_d$, and a lesion age loss $\mathcal{L}_a$.

\begin{align*}
	\mathcal{L} =
	\sum_{i=1}^{N}(
	\lambda_p\mathcal{L}_p +
	\mathbbm{1}_{\{p_i\neq\emptyset\}}(
	 & \lambda_b\mathcal{L}_b + \lambda_g\mathcal{L}_g +                        \\[-1ex]
	 & \lambda_a\mathcal{L}_a + \lambda_f\mathcal{L}_f + \lambda_d\mathcal{L}_d
	)
	)
	\numberthis
\end{align*}

We used cross-entropy for the lesion loss $\mathcal{L}_p$. For the bounding box losses, L1 loss $\mathcal{L}_b$ and the generalized intersection over union \cite{rezatofighi2019generalized} $\mathcal{L}_g$ were used. The segmentation losses comprised of Focal loss $\mathcal{L}_f$ with $\alpha = 0.25$ and $\gamma = 2$ as recommended by Lin \textit{et al.} \cite{lin2017focal}, and Dice loss \cite{milletari2016v} $\mathcal{L}_d$. To enable the uncertainty of lesion age estimates to be quantized, we used quantile loss for the lesion age loss $\mathcal{L}_a$. We predict three quantiles, assuming that estimates for lesion age are normally distributed, that would correspond to minus one standard deviation from the mean, the mean, and plus one standard deviation from the mean. These can be calculated using $\phi$, the cumulative distribution function (CDF) of the standard normal distribution: $\tau_1 = \phi(-1) \approx 0.159$; $\tau_2 = 0.5$; $\tau_3 = \phi(1) \approx 0.841$.

\begin{align*}
	\mathcal{L}_a(a_{\sigma(i)}, \hat{a}_{i}) =
	\sum_{j=1}^3
	\max(
	 & \tau_j||a_{\sigma(i)} - \hat{a}_{i,j}||_1,    \\[-1ex]
	 & (1-\tau_j)||a_{\sigma(i)} - \hat{a}_{i,j}||_1
	)
	\numberthis
\end{align*}

\noindent In order to account for the varying difficulties of each task common to MTL procedures, we employ a random-weighted loss function where weights are drawn from the Dirichlet distribution \cite{lin2021closer}.

\begin{align*}
	\lambda_p,\lambda_b,\lambda_g,\lambda_a,\lambda_f,\lambda_d \overset{\text{i.i.d.}}{\sim} \text{Dir}(1,1,1,1,1,1)
	\numberthis
\end{align*}

\subsection{Inference}

At inference time, we combine lesion age estimates if their associated predicted segmentation masks are connected in 3D. Given a set of $K$ lesion age quantile predictions $\hat{a} = \{\hat{a}_{k}\}_{k=1}^{K}, \hat{a}_{k} \in \mathbb{R}^3$, we estimate probability density functions (PDF) using $f(x;\mu,\sigma_1,\sigma_2)$, the split normal distribution PDF, where $\mu_k=\hat{a}_{k,2}$, $\sigma_{k,1}=\hat{a}_{k,2}-\hat{a}_{k,1}$, and $\sigma_{k,2}=\hat{a}_{k,3}-\hat{a}_{k,2}$ for each instance.

\begin{align*}
	\label{eq:split-normal-pdf}
	f(x;\mu,\sigma_1,\sigma_2) & =
	\begin{cases}
		A \exp \left(- \frac {(x-\mu)^2}{2 \sigma_1^2}\right) & x < \mu   \\
		A \exp \left(- \frac {(x-\mu)^2}{2 \sigma_2^2}\right) & x \ge \mu
	\end{cases} ,        \\[1ex]
	\text{where} \  A          & =\sqrt{2/\pi} (\sigma_1+\sigma_2)^{-1}
	\numberthis
\end{align*}

The maximum argument of the sum of these probability density functions is then the combined lesion age estimate, $\hat{a}_{\mu}$. In the rare instances where a set of predictions produces a negative $\sigma_{k,1}$ or $\sigma_{k,2}$, we resort to the mean lesion age estimate, $\bar{\mu}_{k}$.

\begin{align*}
	\hat{a}_{\mu} = \underset{x}{\mathrm{argmax}}\sum_{k=1}^{K}f(x;\mu_k;\sigma_{k,1};\sigma_{k,2})
	\numberthis
\end{align*}

\section{Experiments}

\subsection{Materials}

\subsubsection{Dataset}

Experiments were conducted on a dataset of 776 acute stroke patients with a known time of onset collected across two clinical sites from 2013 to 2019. Extraction and anonymisation of the images followed the pipeline recommended by Muschelli \cite{muschelli2019recommendations}. The median image size was 512 × 512 × 187 voxels with a spatial resolution of 0.45mm × 0.45mm × 0.8mm. Ground truth segmentation masks of 79,959 slices were produced by manual annotation from experts. Lesion ages were calculated using the time from symptom onset to imaging and log-transformed to account for skewed distribution. Patients were randomly divided such that a fixed 20\% split were used for testing and the remainder for training and validation using a five-fold group cross-validation approach. \autoref{tab:dataset-demographics} lists the characteristics of these groups. When optimizing hyperparameters, 20\% of the total dataset was used for validation. An additional independent dataset of 150 patients was collected from the same clinical sites using a similar methodology in order to validate lesion age estimation performance. Instead of producing segmentation masks, experts selected a total of 4,951 lesion containing slices. Full ethical approval was granted by Wales REC 3 reference number 16/WA/0361.

\begin{table}
	\caption{Population characteristics of the clinical dataset}
	\label{tab:dataset-demographics}
	\centering
	\begin{tabular*}{\linewidth}{@{\extracolsep{\fill}}m{3.1cm}cc}
		\toprule
		\multicolumn{1}{c}{Characteristic} & \multicolumn{1}{m{2.75cm}}{\centering Train and validation set (n = 627)} &
		\multicolumn{1}{m{1.75cm}}{\centering Test set \\ (n = 149)} \\
		\midrule
		Age (years), median (IQR) & 74.9 (63.9-82.8) & 74.7 (63.1-83.0) \\
		\addlinespace[0.25em]
		% Male sex, n (\%) & 317 (50.6\%) & 71 (47.7\%) \\
		Sex, n (\%) & & \\
		\quad Male & 317 (50.6\%) & 71 (47.7\%) \\
		\quad Female & 302 (48.2\%) & 74 (49.7\%) \\
		\quad Missing & 8 (1.3\%) & 4 (2.7\%) \\
		\addlinespace[0.25em]
		ASPECTS, median (IQR) & 9 (8-10) & 9 (8-10) \\
		\addlinespace[0.25em]
		NIHSS on admission, median (IQR) & 13 (7-20) & 13 (7-19) \\
		\addlinespace[0.25em]
		Affected side, n (\%) & & \\
		\quad Left & 335 (53.4\%) & 88 (59.1\%) \\
		\quad Right & 292 (46.6\%) & 61 (40.9\%) \\
		\addlinespace[0.25em]
		Time from symptom onset to CT (minutes), median (IQR) & 232 (109-1212) & 253 (110-1325) \\
		\bottomrule
	\end{tabular*}
	\begin{tabular}{p{\dimexpr\linewidth-2\tabcolsep\relax}}
		\addlinespace[0.5em]
		\begin{footnotesize}
			IQR = Interquartile range; ASPECTS = Alberta stroke programme early CT score \cite{barber2000validity}; NIHSS = National Institutes of Health Stroke Scale \cite{brott1989measurements}
		\end{footnotesize}
	\end{tabular}
\end{table}

\subsubsection{Evaluation}

To evaluate lesion segmentation, we compared mean DSC and intersection over union (IOU) between model predictions and expert segmentation's on a per-subject level. The Mann-Whitney U test was used to determine significance. Identifying the presence of a lesion is more clinically useful than perfect segmentation therefore, we also computed the lesion detection accuracy (LD-ACC), the percentage of expert segmentations $s_{i,j}$ that overlapped with predictions $\hat{s}_{i,j}$ where $i$ and $j$ represent indices for the lesion and image slice, respectively.

\begin{align*}
	\label{eq:lesion-detection-accuracy}
	 & \text{LD-ACC} = \frac{1}{I} \sum_{i=1}^I neq(\sum_{j=1}^J s_{i,j} \hat{s}_{i,j}) \\
	 & \ \ \text{where } neq(x) =
	\begin{cases}
		0 & x = 0    \\
		1 & x \neq 0
	\end{cases}
	\numberthis
\end{align*}

For lesion age, we excluded subjects with lesions of different ages and calculated the coefficient of determination (R$^2$), mean absolute error (MAE), and root mean squared error (RMSE). We also evaluated the classification of lesion age within 4.5 hours of onset using accuracy (ACC) and AUC. These were then computed for the regression models by arranging the $I$ total lesions such that $i = 1,...,I_1$ lesions had an age less than 4.5 hours and $i = I_1 + 1,...,I$ lesions had an age greater than 4.5 hours, where ${a}_{i}$ represents the real and $\hat{a}_{i}$ the predicted age.

\begin{align*}
	\label{eq:lesion-classification-acc}
	 & \text{ACC} = \frac {1}{I_1} \sum_{i=1}^{I_1} acute(\hat{a}_{i}) + \frac {1}{I - I_1} \sum_{j=I_1}^I 1 - acute(\hat{a}_{i}) \\
	 & \ \ \text{where } acute(x) =
	\begin{cases}
		0 & x >             \\
		1 & x \leq \log 270
	\end{cases}
	\numberthis
\end{align*}

\begin{align*}
	\label{eq:lesion-classification-auc}
	 & \text{AUC} = \frac {\sum_{i=1}^{I_1} \sum_{j=I_1}^I S(\hat{a}_{i}, \hat{a}_{j})}{I_1 (I - I_1)} \\
	 & \ \ \text{where } S(x, y) =
	\begin{cases}
		1   & x > y \\
		0.5 & x = y \\
		0   & x < y
	\end{cases}
	\numberthis
\end{align*}

\subsubsection{Implementation}

All models were implemented using PyTorch \cite{NEURIPS2019_9015} version 1.10 and trained from scratch for 100 epochs on a computer with 3.80GHz Intel\textsuperscript{\textregistered} Core\textsuperscript{TM} i7-10700K CPU and an NVIDIA GeForce RTX 3080 10GB GPU. The AdamW \cite{loshchilov2017decoupled} optimizer was used with a weight decay of $10^{-4}$. Learning rate was adjusted from $10^{-6}$ to $10^{-2}$ per-epoch using a cyclical schedule \cite{smith2017cyclical} and exponentially decayed per-cycle with $\gamma=0.92$. Gradient clipping \cite{mikolov2012statistical} was applied to ensure a maximal gradient norm of 0.1. We also employed the stochastic weight averaging \cite{izmailov2018averaging} for the last 5 cycles. During training, lesion containing regions were linearly sampled from the original volumes to a uniform size, 512 × 512 × 1 for 2D and 128 × 128 × 48 for 3D models, with a spatial resolution of 0.45mm × 0.45mm × 0.8mm. The same CT-specific augmentation strategy was applied for all models. Pixel intensities were clipped based on the 0.5 and 99.5th percentile then normalized using Z-score. Inference of the proposed model required about 14 seconds per subject.

\subsection{Results}

\subsubsection{Comparison with Baseline}

\begin{table*}
	\caption{Lesion age estimation and segmentation (mean $\pm$ standard deviation) results obtained by our method and ablation variants compared to the single-task baseline models}
	\label{tab:results-single}
	\centering
	\begin{tabular*}{0.9\linewidth}{@{\extracolsep{\fill}}rcccccccccc}
		\toprule
		& & & \multicolumn{3}{c}{Regression} & \multicolumn{2}{c}{Classification} & \multicolumn{3}{c}{Segmentation} \\
		\cmidrule(lr){4-6}
		\cmidrule(lr){7-8}
		\cmidrule(lr){9-11}
		\multicolumn{1}{c}{Model} & Size & Flops & R$^2$ & MAE & RMSE & AUC & ACC & DSC & IOU & LD-ACC \\
		\midrule
		Intensity GLM & 2 & 2 & 0.365 & 0.816 & 1.021 & 0.858 & 79.5 & \NA & \NA & \NA \\
		ResNet-50 & 24M & 21G & 0.308 & 0.862 & 1.115 & 0.906 & 83.1 & \NA & \NA & \NA \\
		ResNeXt-50 & 23M & 22G & 0.402 & 0.800 & 1.037 & 0.908 & 86.5 & \NA & \NA & \NA \\
		ConvNeXt-T & 28M & 23G & 0.392 & 0.812 & 1.011 & 0.905 & 86.0 & \NA & \NA & \NA \\
		Ours & 40M & 30G & \bfseries 0.513 & 0.680 & \bfseries 0.935 & \bfseries 0.933 & \bfseries 88.5 & 38.2 $\pm$ 24.2 & \bfseries 26.6 $\pm$ 21.0 & \bfseries 98.0 \\
		2D U-Net & 8M & 48G & \NA & \NA & \NA & \NA & \NA & 35.3 $\pm$ 30.0 & 26.2 $\pm$ 26.2 & 95.3 \\
		3D U-Net & 39M & ~91G\dag & \NA & \NA & \NA & \NA & \NA & 36.7 $\pm$ 28.2 & 26.4 $\pm$ 26.4 & 97.3 \\
		TransUNet & 108M & 169G & \NA & \NA & \NA & \NA & \NA & 36.9 $\pm$ 27.7 & 26.1 $\pm$ 25.9 & 97.3 \\
		($P$-value)* & & & & & & & & (0.038) & (0.049) & \\
		\midrule
		ResNet-50 ($\text{L}_1$) & 24M & 21G & 0.297 & 0.866 & 1.124 & 0.904 & 81.7 & \NA & \NA & \NA \\
		ResNeXt-50 ($\text{L}_1$) & 23M & 22G & 0.396 & 0.809 & 1.112 & 0.907 & 84.5 & \NA & \NA & \NA \\
		ConvNeXt-T ($\text{L}_1$) & 28M & 23G & 0.388 & 0.824 & 1.066 & 0.902 & 85.3 & \NA & \NA & \NA \\
		Ours ($\text{L}_1$) & 40M & 30G & 0.503 & \bfseries 0.636 & 0.944 & 0.912 & 86.5 & \bfseries 38.2 $\pm$ 24.1 & 26.3 $\pm$ 21.1 & 98.0\\
		Ours (only segmentation) & 40M & 30G & \NA & \NA & \NA & \NA & \NA & 37.2 $\pm$ 25.1 & 25.3 $\pm$ 24.2 & 97.3 \\
		Ours (only age estimation) & 40M & 30G & 0.510 & 0.682 & 0.938 & 0.930 & 86.8 & \NA & \NA & \NA \\
		Ours (no PPM) & 30M & 28G & 0.330 & 0.733 & 1.097 & 0.874 & 79.7 & 36.0 $\pm$ 24.0 & 24.9 $\pm$ 20.8 & 96.6 \\
		Ours (no GPSA) & 40M & 30G & 0.449 & 0.664 & 0.995 & 0.913 & 83.8 & 35.4 $\pm$ 24.6 & 24.9 $\pm$ 20.7 & 95.3 \\
		Ours (no RLW) & 40M & 30G & 0.402 & 0.675 & 1.036 & 0.904 & 83.4 & 35.0 $\pm$ 25.0 & 24.5 $\pm$ 21.5 & 95.3 \\
		Ours (no DA) & 40M & 30G & 0.025 & 0.945 & 1.357 & 0.756 & 71.6 & 31.6 $\pm$ 24.6 & 21.7 $\pm$ 20.6 & 94.6 \\
		\bottomrule
	\end{tabular*}
	\begin{tabular}{p{\dimexpr0.9\linewidth-2\tabcolsep\relax}}
		\addlinespace[0.5em]
		\begin{footnotesize}
			MAE = Mean absolute error; RMSE = root mean squared error; AUC = Area under the receiver operator characteristic curve; ACC = Accuracy; DSC = Dice similarity coefficient; IOU = intersection over union; LD-ACC = Lesion detection accuracy; GLM = Generalized linear model; $\text{L}_1$ = L1 loss; PPM = Pyramid pooling module; GPSA = Gated positional self-attention; RLW = Random loss weighting; DA = Data augmentation
		\end{footnotesize}
	\end{tabular}
	\begin{tabular}{p{\dimexpr0.9\linewidth-2\tabcolsep\relax}}
		\addlinespace[0.33em]
		\begin{footnotesize}
			*$P$-values are between the results of the proposed model and the next best competing model
		\end{footnotesize} \\
		\begin{footnotesize}
			\dag Normalized by total flops per subject divided by number of slices
		\end{footnotesize}
	\end{tabular}
\end{table*}

\begin{figure*}
	\centering
	\includegraphics[width=0.8\textwidth]{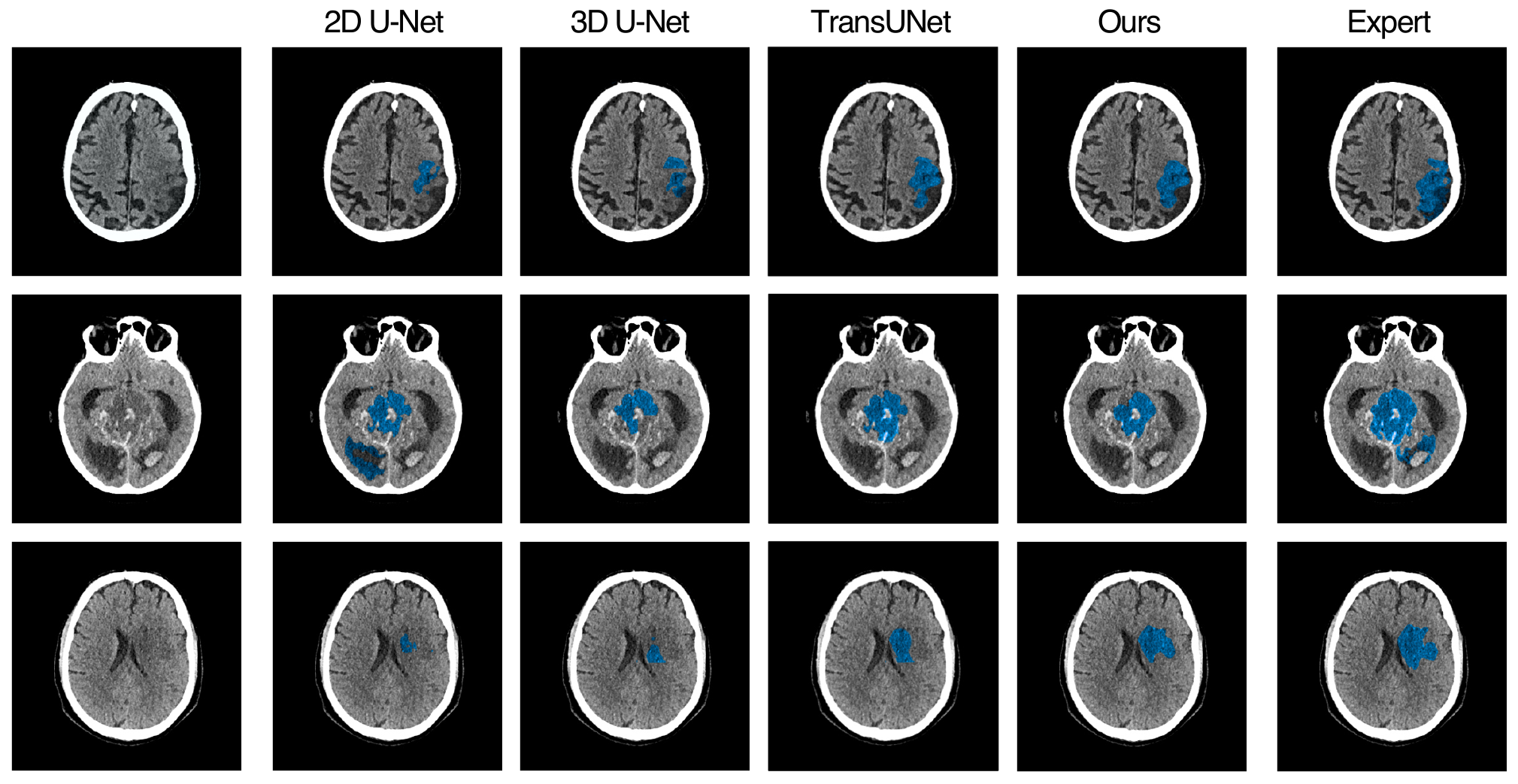}
	\caption{Example lesion segmentations of our method compared to the single-task baseline models.}
	\label{fig:segmentation-results}
\end{figure*}

\begin{figure}
	\centering
	\includegraphics[width=0.8\columnwidth]{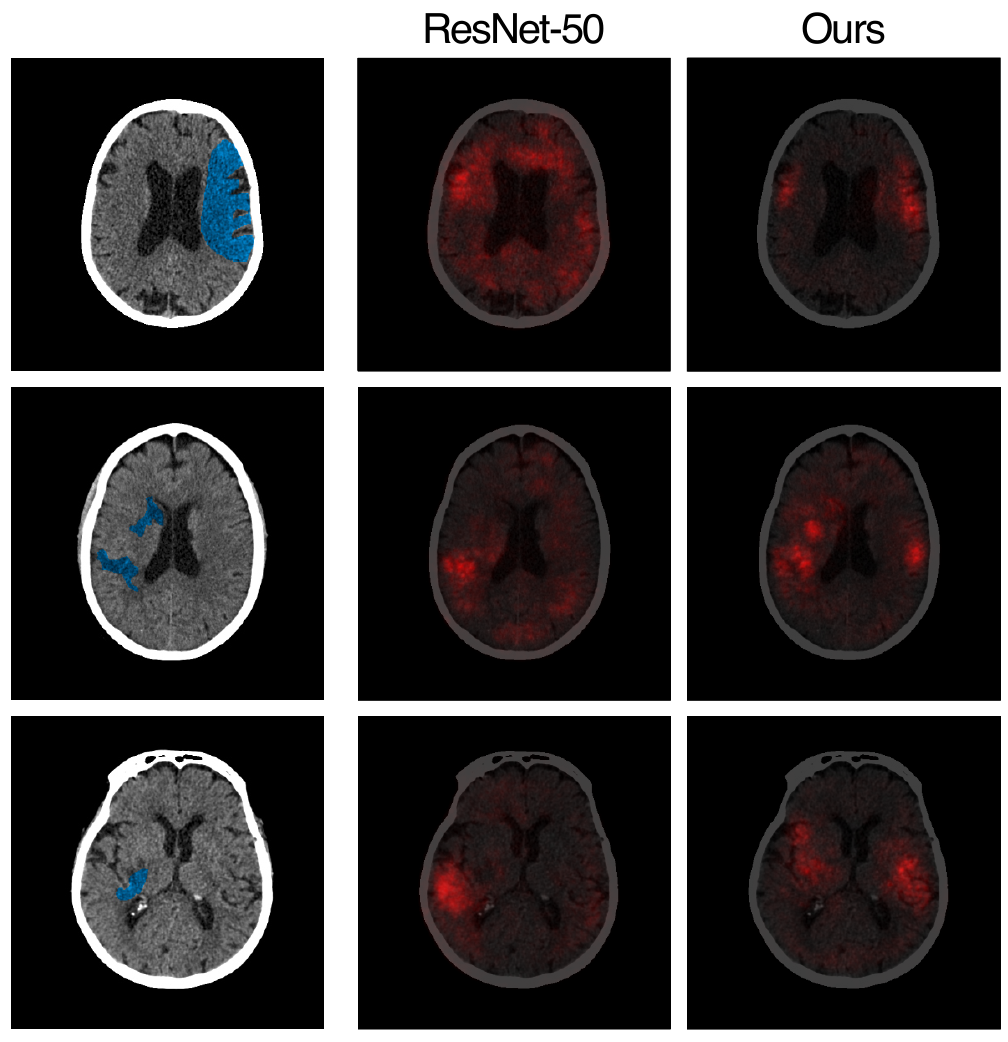}
	\caption{Saliency maps for estimating lesion age of the proposed method compared to ResNet-50. The proposed method has more focused attention on the lesion and appears to have learned to utilize brain from the unaffected corresponding side.}
	\label{fig:regression-saliency}
\end{figure}

We first compare our proposed model to task-specific deep-learning algorithms due to the absence of established methods to jointly perform segmentation and regression. The quantitative results are shown in \autoref{tab:results-single}. For segmentation, we compare against 2D \cite{ronneberger2015u}, 3D U-Net \cite{isensee2021nnu}, and TransUNet \cite{chen2021transunet} using the same Focal and Dice loss function. In this task, our proposed model performs slightly better, with significant ($p$ value $\leq$0.05) increases in DSC and IOU at the expense of generally greater computational demands. While the metrics may seem low, it is unlikely to preclude clinical utility due to the high lesion detection accuracies (LDD-ACC), with only 2\% of lesions going undetected. Notably, despite the proposed model being 2D in nature, it performed competitively against 3D U-Net, suggesting that for lesion segmentation, the ability to capture global semantic information may outweigh the benefits of learning volumetric relations. These findings are also supported by qualitative evaluation as seen in \autoref{fig:segmentation-results}.

For lesion age estimation, we first trained a linear model based on intensity using a similar methodology to Mair \textit{et al.} \cite{mair2021feasibility}. We also trained ResNet-50 \cite{he2016deep}, ResNeXt-50-32x4d \cite{xie2017aggregated}, and ConvNeXt-T \cite{liu2022convnet} models using the same quantile loss function. Compared to these models, our proposed method outperforms them by large margins for all metrics tested. It seems, therefore, that explicit supervised learning of both tasks may be mutually beneficial and is particularly useful in estimating lesion age.

To better understand how the proposed model estimates lesion age, we produce saliency maps for test images by backpropagating back to the input \cite{simonyan2013deep}. As seen in \autoref{fig:regression-saliency}, compared to ResNet-50, the proposed method has generally more focused attention on the lesion. Interestingly, and similar to the aforementioned intensity approach, the proposed model also appears to be utilizing brain in the corresponding spatially distant unaffected side. This perhaps reinforces the benefit of the model's wide-receptive field afforded to it by the use of a Transformer.

\subsubsection{Comparison with other Multi-Task Models}

To further explore the synergy between tasks and the effectiveness of the proposed method, we also compare it with recent multi-task learning networks shown in \autoref{tab:results-multi}. These include 3D variants of MA-MTLN\cite{zhang20213d} and C\textsubscript{MS}VNet\textsubscript{Iter}\cite{zhou2021multi}, which are capable of joint segmentation and classification though not regression of medical images. For both lesion age estimation and segmentation, we find that the multi-task models perform equivalently or are superior to the single-task baseline models. We also note that the proposed method achieves the best performance of the multi-task models and has a greater lead in lesion classification than segmentation. It's possible that this disparity, favoring classification, may be the result of the proposed model being able to utilize continuous lesion ages, whereas the other models were architecturally limited to binary labels.

\begin{figure}
	\centering
	\includegraphics[width=0.8\columnwidth]{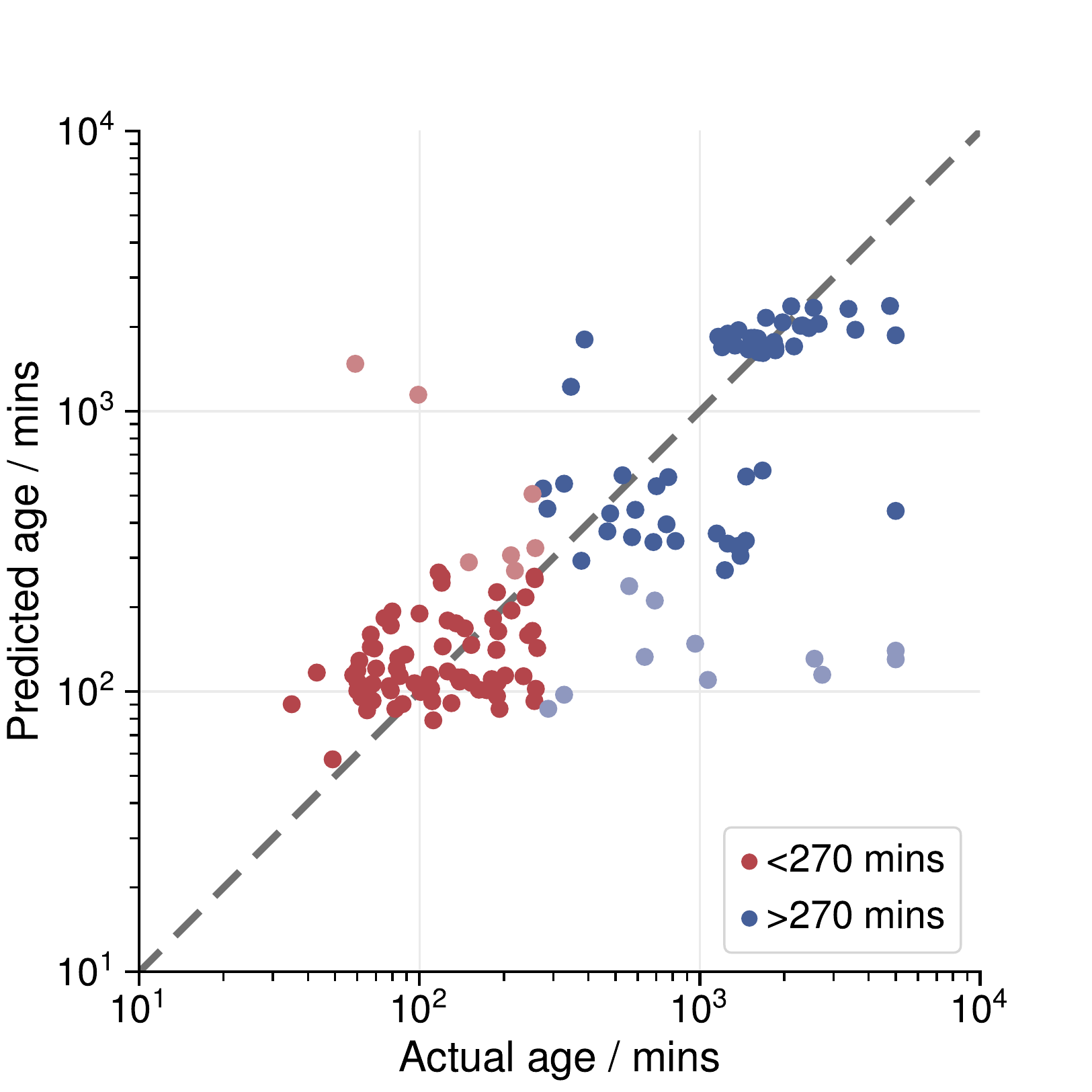}
	\caption{Scatter plot of predicted versus actual lesion ages for the proposed model on the test set.}
	\label{fig:regression-performance}
\end{figure}

\begin{figure}
	\centering
	\includegraphics[width=0.86\columnwidth]{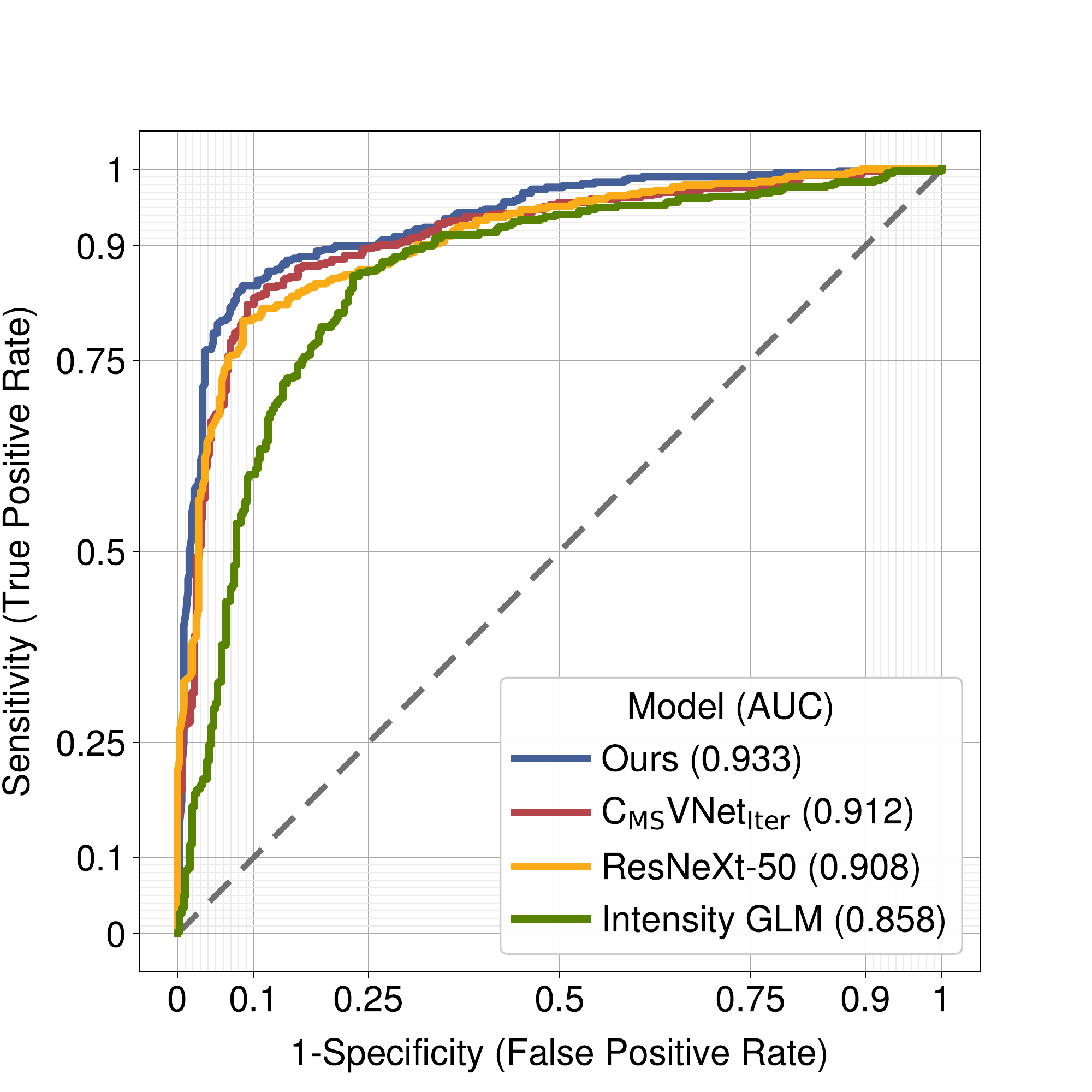}
	\caption{Receiver operating characteristic (ROC) curves of our method compared to the best competing single- and multi-task baselines for classifying lesions ages.}
	\label{fig:roc-performance}
\end{figure}

\begin{table*}
	\caption{Lesion segmentation and classification results (mean $\pm$ standard deviation) obtained by our method compared to multi-task models}
	\label{tab:results-multi}
	\centering
	\begin{tabular*}{0.8\linewidth}{@{\extracolsep{\fill}}rccccccc}
		\toprule
		& & & \multicolumn{2}{c}{Classification} & \multicolumn{3}{c}{Segmentation} \\
		\cmidrule(lr){4-5}
		\cmidrule(lr){6-8}
		\multicolumn{1}{c}{Model} & Size & Flops & AUC & ACC & DSC & IOU & LD-ACC \\
		\midrule
		MA-MTLN & 10M & 30G\dag & 0.907 & 86.0 & 37.0 $\pm$ 23.2 & 26.4 $\pm$ 26.2 & 96.6 \\
		C\textsubscript{MS}VNet\textsubscript{Iter} & 92M & 95G\dag & 0.912 & 86.5 & 37.4 $\pm$ 26.4 & \bfseries 26.7 $\pm$ 24.3 & 97.3 \\
		Ours & 40M & 30G & \bfseries 0.933 & \bfseries 88.5 & \bfseries 38.2 $\pm$ 24.2 & 26.6 $\pm$ 21.0 & \bfseries 98.0 \\
		($P$-value)* & & & & & (0.047) & (0.449) & \\
		\bottomrule
	\end{tabular*}
	\begin{tabular}{p{\dimexpr0.9\linewidth-2\tabcolsep\relax}}
		\addlinespace[0.5em]
		\begin{footnotesize}
			AUC = Area under the receiver operator characteristic curve; ACC = Accuracy; DSC = Dice similarity coefficient; IOU = intersection over union; LD-ACC = Lesion detection accuracy
		\end{footnotesize}
	\end{tabular}
	\begin{tabular}{p{\dimexpr0.9\linewidth-2\tabcolsep\relax}}
		\addlinespace[0.33em]
		\begin{footnotesize}
			*$P$-values are between the results of the proposed model and the next best competing model
		\end{footnotesize} \\
		\begin{footnotesize}
			\dag Normalized by total flops per subject divided by number of slices
		\end{footnotesize}
	\end{tabular}
\end{table*}

\subsubsection{Comparison with the State-of-the-Art}

There are few works that we can compare our results. For segmentation, we are aware of only two studies \cite{barros2020automated,el2021evaluating} that used ground truth NCCT annotations. As argued by El-Hariri \textit{et al.} \cite{el2021evaluating}, direct comparison with studies using annotations from other modalities such as MRI is hindered by the different underlying physiological processes which lead to visible changes. Compared with these studies, the proposed model performs slightly better on this challenging task with a DSC of 38.2\% compared to 37\% by Barros \textit{et al.} \cite{barros2020automated} and 37.7\% by El-Hairi \textit{et al.} \cite{el2021evaluating}. For lesion age estimation, the proposed model achieved an AUC of 0.933 for classifying whether a stroke event is within 4.5 hours of onset. Similar to the predominately manual methods by Broocks \textit{et al.} \cite{broocks2020lesion} and Mair \textit{et al.} \cite{mair2021feasibility} with reported AUC of 0.91 and 0.955, respectively. However, we note that due to the dynamic nature of ischemia, the classification of older lesions is considerably easier. This is noticeable in \autoref{fig:regression-performance} where the proposed model predictions showed better agreement with lesions of a greater age. Therefore, the difficulty of this task is highly dependent on the distribution of lesion ages in the dataset, and without an open benchmark, objective assessment against other methods is limited. This is further supported by \autoref{fig:roc-performance} where our intensity model achieves an AUC of only 0.858 using a similar methodology to these studies. Performance aside and in contrast to prior works, the proposed approach benefits from being fully automated and naturally able to accommodate patients with multiple lesions.

\subsubsection{Generalization to Other Datasets}

\begin{table}
	\caption{Lesion age estimation results obtained by our method compared to the single- and multi-task models for a second independent test set}
	\label{tab:results-test}
	\centering
	\begin{tabular*}{\linewidth}{@{\extracolsep{\fill}}rccccc}
		\toprule
		& \multicolumn{3}{c}{Regression} & \multicolumn{2}{c}{Classification} \\
		\cmidrule(lr){2-4}
		\cmidrule(lr){5-6}
		\multicolumn{1}{c}{Model} & R$^2$ & MAE & RMSE & AUC & ACC \\
		\midrule
		Intensity GLM & 0.124 & 0.920 & 1.331 & 0.772 & 78.0 \\
		ResNet-50 & 0.280 & 0.861 & 1.102 & 0.892 & 83.3 \\
		ResNeXt-50 & 0.382 & 0.795 & 1.043 & 0.904 & 86.7 \\
		ConvNeXt-T & 0.385 & 0.804 & 1.031 & 0.905 & 86.0 \\
		Ours & \bfseries 0.511 & \bfseries 0.682 & \bfseries 0.958 & \bfseries 0.921 & \bfseries 88.7 \\
		MA-MTLN & \NA & \NA & \NA & 0.911 & 86.7 \\
		C\textsubscript{MS}VNet\textsubscript{Iter} & \NA & \NA & \NA & 0.907 & 87.3 \\
		\bottomrule
	\end{tabular*}
	\begin{tabular}{p{\dimexpr\linewidth-2\tabcolsep\relax}}
		\addlinespace[0.5em]
		\begin{footnotesize}
			MAE = Mean absolute error; RMSE = root mean squared error; AUC = Area under the receiver operator characteristic curve; ACC = Accuracy
		\end{footnotesize}
	\end{tabular}
\end{table}

\begin{table}
	\caption{Lesion segmentation results (mean $\pm$ standard deviation) obtained by our method compared to the single- and multi-task models for the ISLES-2018 dataset}
	\label{tab:results-isles}
	\centering
	\begin{tabular*}{\linewidth}{@{\extracolsep{\fill}}rccc}
		\toprule
		\multicolumn{1}{c}{Model} & DSC & IOU & LD-ACC \\
		\midrule
		2D U-Net & 18.3 $\pm$ 12.9 & 10.2 $\pm$ 9.3 & 73.0 \\
		3D U-Net & 11.1 $\pm$ 13.7 & 6.5 $\pm$ 8.7 & 71.4 \\
		TransUNet & 17.7 $\pm$ 14.6 & 9.8 $\pm$ 10.1 & 71.4 \\
		Ours & \bfseries 20.3 $\pm$ 11.5 & 11.2 $\pm$ 9.1 & \bfseries 74.6 \\
		MA-MTLN & 19.7 $\pm$ 14.1 & \bfseries 11.4 $\pm$ 9.6 & 73.0 \\
		C\textsubscript{MS}VNet\textsubscript{Iter} & 19.8 $\pm$ 13.8 & 11.0 $\pm$ 9.8 & 73.0 \\
		($P$-value)* & (0.187) & (0.356) & \\
		\bottomrule
	\end{tabular*}
	\begin{tabular}{p{\dimexpr\linewidth-2\tabcolsep\relax}}
		\addlinespace[0.5em]
		\begin{footnotesize}
			DSC = Dice similarity coefficient; IOU = intersection over union; LD-ACC = Lesion detection accuracy
		\end{footnotesize}
	\end{tabular}
	\begin{tabular}{p{\dimexpr\linewidth-2\tabcolsep\relax}}
		\addlinespace[0.33em]
		\begin{footnotesize}
			*$P$-values are between the results of the proposed model and the next best competing model
		\end{footnotesize}
	\end{tabular}
\end{table}

To the best of our knowledge, there are no publicly available datasets for ischemic lesion age estimation. For this reason, we utilized an additional independently collected dataset to further validate the performance of our approach with the results shown in \autoref{tab:results-test}. Encouragingly, these findings reveal no unexpected outcomes, with our method performing best in both regression and classification tasks, which aligns with our initial analysis.

For lesion segmentation, while there are no datasets with ground truth NCCT annotations, the ISLES (Ischemic Stroke Lesion Segmentation) challenge \cite{hakim2021predicting} contains images with labels derived from diffusion-weighted imaging. As previously noted, this represents a related, although inherently different, task \cite{el2021evaluating}. Additionally, the population characteristics are markedly different with lower resolution scans and younger patients with higher National Institutes of Health Stroke Scale (NIHSS) scores \cite{hakim2021predicting}. Therefore, we also assess out-of-domain generalization by comparing the performance of the trained proposed model to trained single- and multi-task models on the ISLES 2018 dataset, with results presented in \autoref{tab:results-isles}.

The proposed method achieved the highest DSC and LDD-ACC scores of 20.3\% and 74.6\%, respectively. However, perhaps unsurprisingly, the performance of all models was considerably worse than the best-published approach \cite{song20183d} for this challenge, with a DSC of 51\%. Giving further support to the motivation behind this work, we note that the multi-task models generally outperform the single-task models. Interestingly, it also seems apparent that the 2D models are able to generalize better than the 3D models, which may suggest they are more robust to the differences in slice thickness.

\subsubsection{Ablation Study}

We conducted a series of experiments, shown in \autoref{tab:results-single}, to verify the effectiveness of our method and justify its design decisions. First, we observe that our data augmentation strategy appears to have the largest impact on lesion age estimation and segmentation performance. It seems plausible this may be the result of overfitting from the limited data combined together with the strategy of training from scratch. Second, by jointly training age estimation and segmentation, the performance of both tasks appear to improve modestly. Third, using GPSA, PPM, and RLW rather than equally weighted losses provide benefits primarily to age estimation with comparatively little effect on segmentation. Finally, we note a consistent increase in lesion age estimation performance gained by using our proposed quantile loss based method across all tested models.

\section{Discussion}

Stroke is a leading cause of adult disability and death worldwide. Effective clinical management often relies upon the interpretation of CT imaging to confirm both the occurrence and age of an event. Previous attempts to automate these tasks have treated them independently and thereby may have overlooked their apparent complementary relationship. In the present study, a novel transformer-based approach is proposed to address this that is able to jointly segment and estimate the age of cerebral ischemic lesions. The performance of our method was then characterized through a number of experiments.

\subsection{Limitations}

It should be noted that this study had several limitations. As the methodology relied heavily on supervised machine learning it was, therefore, subject to many general issues and particularly those common to medical image analysis. First, and perhaps the most important, limited sample size due to the amount of data available publicly and the laborious nature of manually annotating additional subjects \cite{balki2019sample}. Second, difficulties in comparing algorithms in an objective manner due to the different populations and evaluation techniques of other published works \cite{kelly2019key}. Third, the introduction of potential bias through the use of data that may not represent the wider population and therefore hinder generalizability. This is of particular importance as all CT images used were obtained from Siemens scanners and previous studies have shown significant variation between scanners and manufacturers \cite{han2006reliability, takao2014effects}. There are also limitations specific to this work. Only one experienced scan reader was used to annotate the images. The labels for lesion age are likely subject to significant noise as they rely upon patients accurately reporting the time of symptom onset. Finally, the performance of the proposed model may be limited due to it's underlying 2D nature, as there are likely aspects that cannot be captured well due to its inability to leverage the whole 3D volume.

\subsection{Future Work}

Subsequent research could seek to address these aforementioned limitations. For example, by extending the proposed method to 3D, which appears relatively straightforward and can be achieved by replacing the backbone CNN encoder and segmentation head with 3D equivalents. However it seems likely that other changes may also be required to maintain computational efficiency. Additionally, before considering real-world deployment, the safety and applicability of our approach must be assessed prospectively in a larger and more extensive clinical study.

\subsection{Broader Impact}

If successfully validated, it is hoped that this work could lead to better outcomes for stroke patients due to faster diagnosis and better choice of treatment. Additionally, by exclusively using NCCT imaging, our method has the potential to be widely applicable, reducing health inequality in areas where medical experts are limited, such as in low- and middle-income countries \cite{wahl2018artificial}. The proposed approach could also find uses outside the intended application, anywhere that concurrent segmentation and regression may be seen as beneficial. For example, within healthcare, to detect a pneumothorax or effusion on chest X-ray and estimate their volumes \cite{hoi2007accurate, brockelsby2016p1}. Or in other domains, such as detecting faces and estimating their age \cite{othmani2020age}.

\section{Conclusion}

In this paper, we proposed a novel transformer-based network for concurrent ischemic lesion segmentation and age estimation of CT brain. By incorporating GPSA layers and using a modality-specific data augmentation strategy, we enhanced the data efficiency of our method. Furthermore, we improved lesion age estimation performance by better combining multiple predictions through the incorporation of uncertainty. Extensive experiments on a clinical dataset demonstrated the effectiveness of our method compared to conventional and task-specific algorithms. Future work includes further prospective clinical validation and exploring the extension of the model to 3D.

\bibliographystyle{IEEEtran}
\bibliography{bibliography}

\end{document}